\documentstyle[11pt]{article}
\title{The influence of correlated hopping on valence and metal-insulator 
transitions in the Falicov-Kimball model}
\author{Pavol Farka\v sovsk\'y\\
Institute  of  Experimental  Physics,  Slovak   Academy   of
Sciences\\
Watsonova 47, 043 53 Ko\v {s}ice, Slovakia}
\date{}
\begin{document}
\baselineskip=22pt
\maketitle

\begin{abstract}
The extrapolation of small-cluster exact-diagonalization
calculations is used to examine the influence of correlated hopping 
on valence and metal-insulator transitions in the one-dimensional 
Falicov-Kimball model. It is shown that the ground-state phase diagram 
as well as the picture of valence and metal-insulator transitions found 
for the conventional Falicov-Kimball model (without correlated hopping)
are strongly changed when the correlated hopping term is added.
The effect of correlated hopping is so strong that it can induce the 
insulator-metal transition, even in the half-filled band case.

\end{abstract}
\vspace{0.5cm}

In the past decade, a considerable amount of effort has  been devoted 
to  understanding  of  the  multitude  of  anomalous physical 
properties of rare-earth and transition-metal compounds.
Recent theoretical works based on exact numerical and analytical calculations 
showed that many of these anomalous features,  e.g., mixed  valence phenomena, 
metal-insulator transitions, etc., can be describe very well within different 
versions of the Falicov-Kimball model~(FKM)~\cite{Falicov}.
For example, it was found that the spinless FKM, in the 
pressure induced case, can describe both types of intermediate-valence
transitions observed experimentally in rare-earth compounds: a discontinuous 
insulator-insulator transition for sufficiently strong 
interactions~\cite{Fark1} and a discontinuous insulator-metal transition 
for weak interactions~\cite{Fark2}. In addition, at nonzero temperatures this 
model is able to provide the qualitative explanation for anomalous large 
values of the specific heat coefficient and for extremely large changes of 
electrical conductivity~\cite{Fark3} found in some intermediate-valence 
compounds (e.g., in SmB$_6$). Moreover, very recently the spin-one-half 
version of the FKM has been used successfully for a description 
of a  discontinuous intermediate-valence  transition (accompanied by 
a discontinuous insulator-metal transition) in SmS~\cite{Fark4} as well as 
for a description of an anomalous magnetic response of the Yb-based 
valence-fluctuating compounds~\cite{Freericks1}. 

These results show that the FKM could, in principle,
yield the correct physics for describing rare-earth compounds.
On the other hand it should be noted that the 
conventional FKM neglects all nonlocal interaction terms and thus it 
is questionable whether above mentioned results persist also in
more realistic situations when nonlocal interactions will be turned 
on. An important nonlocal interaction term obvious absent
in the conventional FKM is the term of correlated hopping
in which the $d$-electron hopping amplitudes $t_{ij}$ between neighbouring 
lattice sites $i$ and $j$ depend explicitly on the occupancy $(f^+_if_i)$ 
of the $f$-electron orbitals. To examine effects of this term on valence 
and metal-insulator transitions in the FKM we adopt here the 
following general form for the nearest-neighbour matrix elements

\begin{equation}
\tilde{t}_{ij}=t_{ij}+t'_{ij}(f^+_if_i+f^+_jf_j),
\end{equation}
which represent a much more realistic type of electron hopping 
than the conventional hopping ($t_{ij}=-t$ if $i$ and $j$ are the 
nearest neighbours, and zero otherwise).

Thus the spinless FKM in which the effects of correlated hopping are
included can be written as
\begin{equation}
H=\sum_{<ij>}t_{ij}d^+_id_j+\sum_{<ij>}t'_{ij}(f^+_if_i+f^+_jf_j)d^+_id_j+
U\sum_if^+_if_id^+_id_i+E_f\sum_if^+_if_i,
\end{equation}
where $f^+_i$, $f_i$ are the creation and annihilation
operators  for an electron in  the localized state at
lattice site $i$ with binding energy $E_f$ and $d^+_i$,
$d_i$ are the creation and annihilation operators
of the itinerant spinless electrons in the $d$-band
Wannier state at site $i$.

The first term of (2) is the kinetic energy corresponding to
quantum mechanical hopping of the itinerant $d$-electrons
between the nearest-neighbour sites $i$ and $j$. 
The second term is just the correlated hopping term discussed above.
The third term is the on-site Coulomb 
interaction between the $d$-band electrons with density
$n_d=\frac{1}{L}\sum_id^+_id_i$ and the localized
$f$-electrons with density $n_f=\frac{1}{L}\sum_iw_i$, where $L$ 
is the number of lattice sites. The last term stands 
for the localized $f$ electrons whose sharp energy level is $E_f$.

Since in this spinless version of the FKM
without hybridization  the $f$-electron occupation
number $f^+_if_i$ of each site $i$ commutes with
the Hamiltonian (2), the $f$-electron occupation number
is a good quantum number, taking only two values: $w_i=1$
or 0, according to whether or not the site $i$ is occupied
by the localized $f$ electron. Therefore the Hamiltonian (2) can 
be written as

\begin{equation}
H=\sum_{<ij>}h_{ij}(w)d^+_id_j+E_f\sum_iw_i,
\end{equation}
where $h_{ij}(w)=\tilde{t}_{ij}(w)+Uw_i\delta_{ij}$ and

\begin{equation}
\tilde{t}_{ij}(w)=t_{ij}+t'_{ij}(w_i+w_j).
\end{equation}

Thus for a given $f$-electron configuration
$w=\{w_1,w_2 \dots w_L\}$ defined on a one-di\-men\-sional
lattice with periodic boundary conditions, the Hamiltonian (2)
is the second-quantized version of the single-particle
Hamiltonian $h(w)$, so the investigation of
the model (3) is reduced to the investigation of the
spectrum of $h$ for different configurations of $f$ electrons.
Since the $d$ electrons do not interact among themselves, the
numerical calculations precede directly in the following steps
(in the next we consider only the case $N_f+N_d=L$,
which is the point of the special interest for the
mixed-valence phenomena).
(i) Having $w=\{w_1,w_2 \dots w_L\}$, $U$, $E_f$ and the nearest-neighbour
hopping amplitudes $t$ and $t'$ fixed, (in the following $t=-1$ and all 
energies are measured in units of $t$) find all eigenvalues 
$\lambda_k$ of $h(w)$. (ii) For a given
$N_f=\sum_iw_i$ determine the ground-state energy
$E(w,U,E_f)=\sum_{k=1}^{L-N_f}\lambda_k+E_fN_f$ of a particular
$f$-electron configuration $w$ by filling in the lowest
$N_d=L-N_f$ one-electron levels. (iii) Find the $w^0$ for which
$E(w,U,E_f)$ has a minimum. Repeating this procedure for different
values of $U,t'$ and $E_f$, one can study directly the 
the ground-state phase diagram and  valence 
transitions (a dependence of the $f$-electron occupation number on the 
$f$-level position $E_f$) in the FKM with correlated
hopping.

The most interesting question that arises for the FKM with 
correlated hopping is whether the correlated hopping term can change the 
ground state phase diagram and the picture of valence and metal-insulator 
transitions found for the conventional FKM  ($t=-1$ and $t'=0$). 
The nature of the ground state, its energetic and structural properties and 
the correlation-induced metal-insulator transitions are subjects of special
interest. For the conventional FKM  these problems are well understood at 
least at half-filling ($E_f=0,n_f=n_d=1/2$). In this case
the localized $f$-electrons fill up one of two sublattices of 
the hypercubic lattice (the charge density wave state) 
and the corresponding ground state is insulating for all $U>0$. 
Thus, for the finite interaction strength there is no correlation-induced 
metal-insulator transition in the half-filled band case. 
Outside half-filling the situation is more complicated since there exist
the analytical results only for a restricted class of 
configurations~\cite{Gruber,Freericks2} and exact-numerical results are
limited to small finite clusters~\cite{Fark1,Fark2}. In spite of these 
restrictions both analytical and numerical results yield the same 
ground-state phase diagram which consists of only two  main domains:
the domain of the phase separation and the domain of the most 
homogeneous configurations. The domain of the phase separation is restricted 
to a small region of low $f$-electron concentrations ($n_f<1/4$) and      
weak Coulomb interactions ($U<1.2$). In this domain the ground state are
incoherent (metallic) mixtures~\cite{Fark2,Gruber} of the empty configuration 
(whose length $l$ is at least $L/2$) and a configuration $w$ 
(whose length is $l < L/2$). Outside this domain the ground states are the 
most homogeneous configurations, which are insulating for all $f$-electron
concentrations. The boundary between these two domains is the boundary
of correlation-induced metal-insulator transitions. 

One can expect, on the base of simple arguments, that the ground-state 
phase diagram of the FKM with correlated hopping will be  fully different 
from one discussed above for the conventional FKM. 
Indeed, the following selection of hopping matrix amplitudes 
$t=-1$ and $t'>0$ may favor the segregated configuration 
$\{11 \dots 100\dots 0\}$ since the itinerant $d$ electrons have the lower 
kinetic energy in this state. This mechanism could lead, for example,  to the
instability of the CDW state $\{10 \dots 10\}$ that is the ground state for 
$t'=0$ and thereby to a metal-insulator transition, even in the 
half-filled band case. 
To verify this conjecture  we have  performed an exhaustive study of the 
model on finite clusters (up to 32 sites) for a wide range of parameters 
$t'$ and $U$. The results of numerical calculations are summarized in 
Fig.~1 in the form of the $t'$-$U$ phase diagram. In addition to the 
alternating configuration $w_{1}=\{10 \dots 10\}$ that is the ground state at 
$t'=0$ for all nonzero $U$ we found two new phases that can be the ground
states of the model, and namely, the alternating configuration 
$w_2=\{1100 \dots 1100\}$ with double period and the segregated configuration 
$w_3=\{11 \dots 100 \dots 0\}$. Thus at nonzero $t'$ the configuration $w_1$ 
becomes unstable against the transition to $w_2$ and $w_3$. The transition
from $w_{1}$ to $w_2$ is the insulator-insulator transition since 
both $w_{1}$ and $w_2$ configurations have the finite gaps~$\Delta$ at the 
Fermi energy~\cite{note} for nonzero values of $U$~(see the inset in Fig.~1). 
On the other hand the segregated configuration $w_3$ is metallic ($\Delta=0$)
for all nonzero $U$ and so the transition from $w_{1}$ to $w_3$ 
(as well as from $w_2$ to $w_3$) is the insulator-metal
transition. Thus we arrive to the very important conclusion, and namely, that 
the correlated hopping term can induce the insulator-metal transition,  even 
in the half-filled band case.
  
The fact that the correlated hopping plays the important role in stabilizing 
the ground state indicates that the picture of valence and metal-insulator 
transitions found in our previous papers within the conventional 
FKM~\cite{Fark1,Fark2} could be dramatically changed if finite values 
of $t'$ will be considered. 
To verify this conjecture we have performed the study of the model for two 
representative values of $U$ ($U=1.5$ and $U=4$) and a selected set of
$t'$ values. The results of numerical calculations obtained for $U=1.5$ 
on finite clusters (up to 32 sites) can be summarized as follows:
(i) For sufficiently small values of $t'$ ($t' \le 0.3 $) the ground states
are the most homogeneous configurations for all $f$-electron 
concentrations. (ii) For intermediate values of $t'$ ($t' \sim 0.6$)
the ground states are the segregated configurations for $n_f<0.5$
and the most homogeneous configurations for $n_f \ge 0.5$.
(iii) For large values of $t'$ ($t'>0.8$) the ground states are 
only the segregated configurations. Since no significant finite-size effects 
have been observed on clusters up to 32 sites we suppose that these
results can be satisfactory extrapolated on large systems and used 
for a study of effects of correlated hopping on valence and metal-insulator 
transitions. Typical valence transitions obtained on the extrapolated set
of configurations are presented in Fig.~2. For $t'=0.3$ the valence 
transition holds all characteristic features found in our 
previous papers for $t'=0$. The basic structure of the transition is formed 
by the most homogeneous configurations with the smallest periods
($n_f=1/5,1/4,1/3,1/2,2/3,3/4,4/5$). 
This primary structure is independent of $L$
while the secondary structure corresponding to remaining configurations 
strongly depends on $L$ and forms the gradual transition between two valence
states from the primary structure. The correlated hopping term reduces 
considerably the width of the primary structure, especially for $n_f \ge 0.5$,
but it does not change the nature of the ground state. The ground states
for $t'=0.3$ are the most homogeneous configurations  which are 
insulating~(see the inset in Fig.~2) and thus there is 
no insulator-metal transition induced by external pressure 
for small values of $t'$. (It is generally supposed that in the pressure 
induced case $E_f$ is proportional to pressure.)    
From this point of view the situation looks more hopeful for intermediate  
values of $t'$. As was mentioned above for this case the ground states are 
the segregated (metallic) configurations (for $n_f < 0.5$) and thus one could
expect the pressure induced insulator-metal transition at $E_f=E_c$ 
(corresponding to $n_f=n_c=0.5$), of course if the segregated configurations 
persist as the ground states on some finite region of $E_f$ values.
Fig.~2 illustrates that such a situation indeed becomes. For $t'=0.6$
the segregated configurations are the ground states in the wide region 
above $E_c=-0.02$ and the valence transition in this region is continuous.
Below $E_c$ the ground states are the most homogeneous (insulating) 
configurations. In this region the primary structure is practically 
fully suppressed and only the intermediate valence states with 
$n_f=1/2$ and $n_f=2/3$ have the finite regions of stability.
At $E_f=E_c$ the FKM exhibits a discontinuous insulator-metal transition
that is apparently induced by correlated hopping. This result shows how 
crucial role plays the correlated hopping in the mechanism of valence and 
insulator-metal transitions. Already relatively small values of $t'$ can
fully destroy the picture of valence and metal-insulator transitions
found for the conventional FKM and therefore the correlated hopping term
cannot be neglected in the correct description of valence and metal-insulator
transitions. Of course, for large values of $t'$ we can expect the largest
deviations from $t'=0$ case, since the ground states in this region 
are only the segregated configurations which contrary to the most homogeneous 
configurations are metallic. The inset in Fig.~2 confirms this conjecture. 
The valence transitions for large values of $t'$ are only continuous and 
fully differ from  $t'=0$ case.

In the strong coupling region ($U=4$) the ground states are either 
the most homogeneous configurations (for small and intermediate  $t'$) 
or the segregated configurations (for large $t'$). 
Thus for small and intermediate values of $t'$ the strong coupling picture 
of valence and metal-insulator transitions will be the same as one for $t'=0$. 
All ground states are insulating and the valence transitions, 
for both conventional FKM and FKM with correlated hopping consist of a few 
discontinuous valence transitions which number is reduced with increasing $U$. 
Only one difference between valence transitions obtained
within these two models is that the width of the valence transition in
the FKM with correlated hopping is reduced due to finite $t'$. 
Fig.~3 illustrates  that the effect is very strong even for relatively small 
values of $t'$. This indicates that the correlated hopping plays important 
role also in the strong coupling limit although unlike the weak-coupling limit
now small values of $t'$ do not change qualitatively the picture of valence 
and metal-insulator transitions. In order to change qualitatively the picture 
of valence and metal-insulator transitions in the strong coupling limit, 
larger values of $t'$ should be considered. For example,
we have found that for $U=4$ and $t'=1.2$ the ground states 
are the segregated (metallic) configurations for all $f$-electron 
concentrations and the corresponding valence transition is continuous. 

The exactly opposite effect on valence transitions in the strong 
coupling limit has the correlated hopping term with negative values of $t'$.
As shown in Fig.~4. the correlated hopping with $t'<0$ stabilizes
the insulating intermediate-valence states corresponding to the most 
homogeneous configurations with the smallest periods 
(which are the ground states in this region) and this effect 
is very strong even for relatively small values of $t'$. 
This again leads to conclusion that the correlated hopping 
must be taken into account in the correct description of valence and 
metal-insulator transitions.

\vspace{0.5cm}
This work was supported by the Slovak Grant Agency VEGA
under grant No. 2/4177/97. Numerical results were obtained using
computational resources of the Computing Centre of the Slovak 
Academy of Sciences.

\newpage
Figure Captions

\vspace{0.5cm}
Fig.~$1.$ 
$t'$-$U$ phase diagram of the FKM with correlated hopping at half-filling
($E_f=0,n_f=n_d=0.5$). Three different
phases correspond to the alternating configuration $w_1=\{10 \dots 10\}$,
the alternating configuration $w_2=\{1100 \dots 1100\}$ with double
period and the segregated configuration $w_3=\{1 \dots 10 \dots 0\}$. 
For $L=24$ the phase diagram has been obtained over the full set
of $f$-electron configurations, while for $L=600$ only the restricted 
set of configurations $w_1, w_2$ and $w_3$ has been used to determine
the phase boundaries. 
The inset shows $t'$-dependence of energy gaps corresponding to the 
ground state configurations for different values of $U$.

\vspace{0.5cm}
Fig.~$2.$ Dependence of the $f$-electron occupation number $n_f$
on the $f$-level position $E_f$ for different values of correlated
hopping $t'$ at~$U=1.5$. 
The upper inset shows energy gaps corresponding to the most homogeneous
configurations that are ground states for $t'=0.3$ (for all $f$-electron
concentrations) and $t'=0.6$ (for $n_f \ge 0.5$).
The lower inset shows examples of $n_f(E_f)$ behaviour for large 
values of $t'$.

\vspace{0.5cm}
Fig.~$3.$ Dependence of the $f$-electron occupation number $n_f$
on the $f$-level position $E_f$ for different values of correlated
hopping $t'$ at~$U=4$.
The inset shows energy gaps corresponding to the most homogeneous
configurations that are ground states for all $f$-electron concentrations
for $t'=0.4$ and $U=4$ 

\vspace{0.5cm}
Fig.~$4.$ Dependence of the $f$-electron occupation number $n_f$
on the $f$-level position $E_f$ for different values of correlated
hopping $t'\le 0$ at~$U=4$.
The inset shows energy gaps corresponding to the most homogeneous
configurations that are ground states for all $f$-electron concentrations
for $t'=-0.2$ and $t'=-0.4$. 


\begin{thebibliography}{99}
\bibitem{note} Since we consider the case $N_f+N_d=L$, 
the Fermi level $E_F$ and the energy gap $\Delta$
of some configuration $w$ are given by $E_F=\lambda_{L-N_f}$ and
$\Delta=\lambda_{L-N_f+1}-\lambda_{L-N_f}$, respectively.
\end{thebibliography}
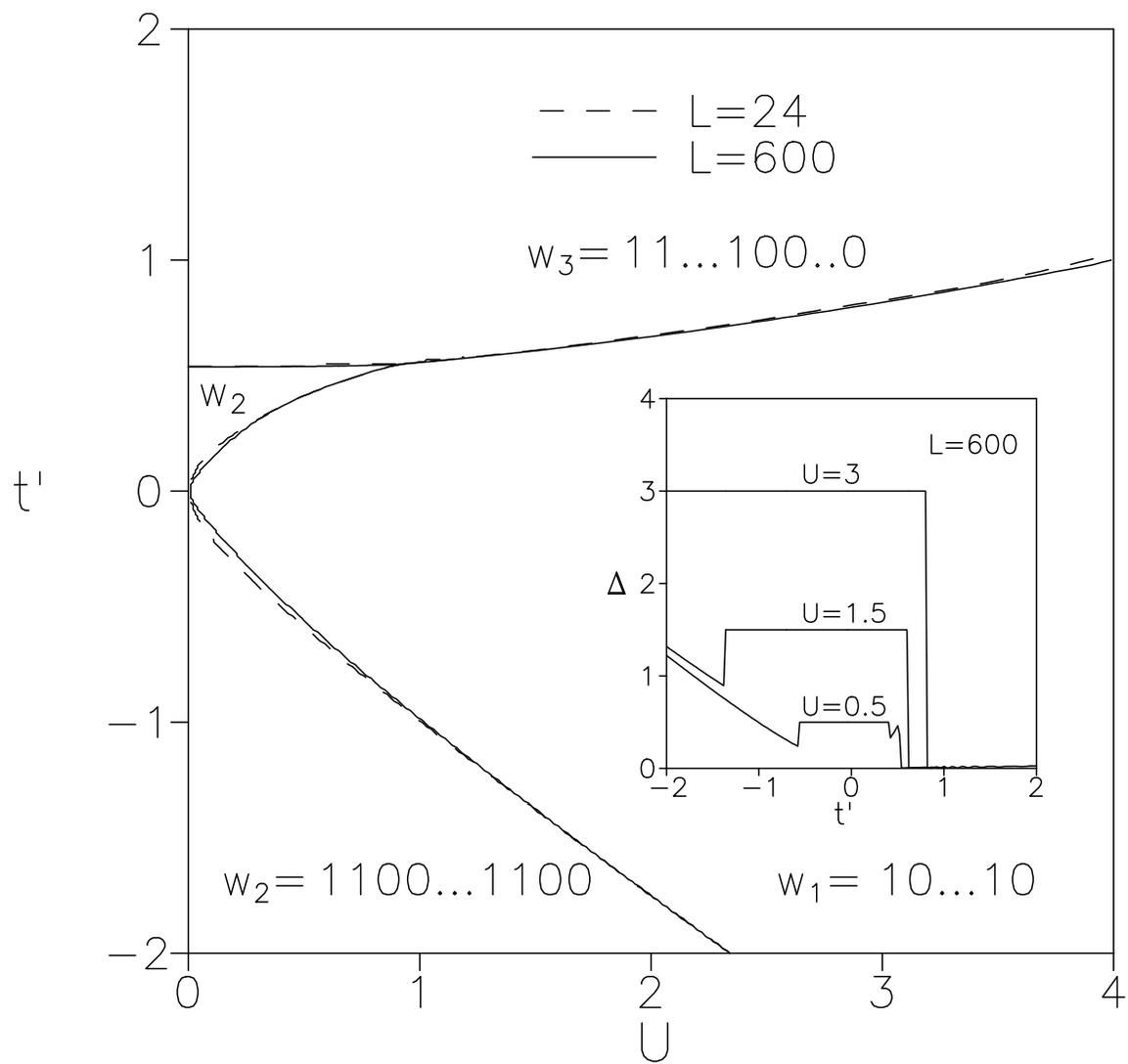
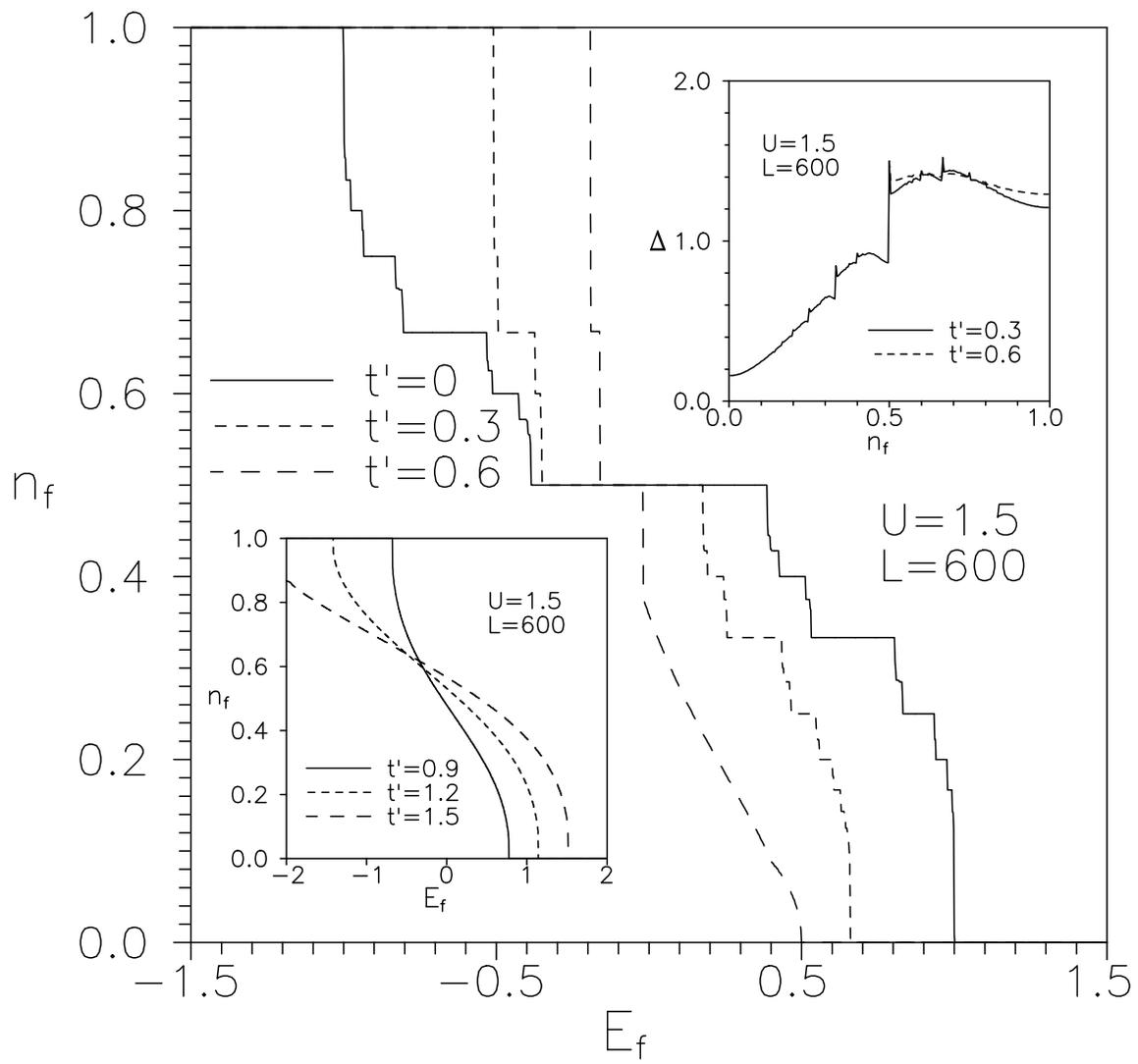
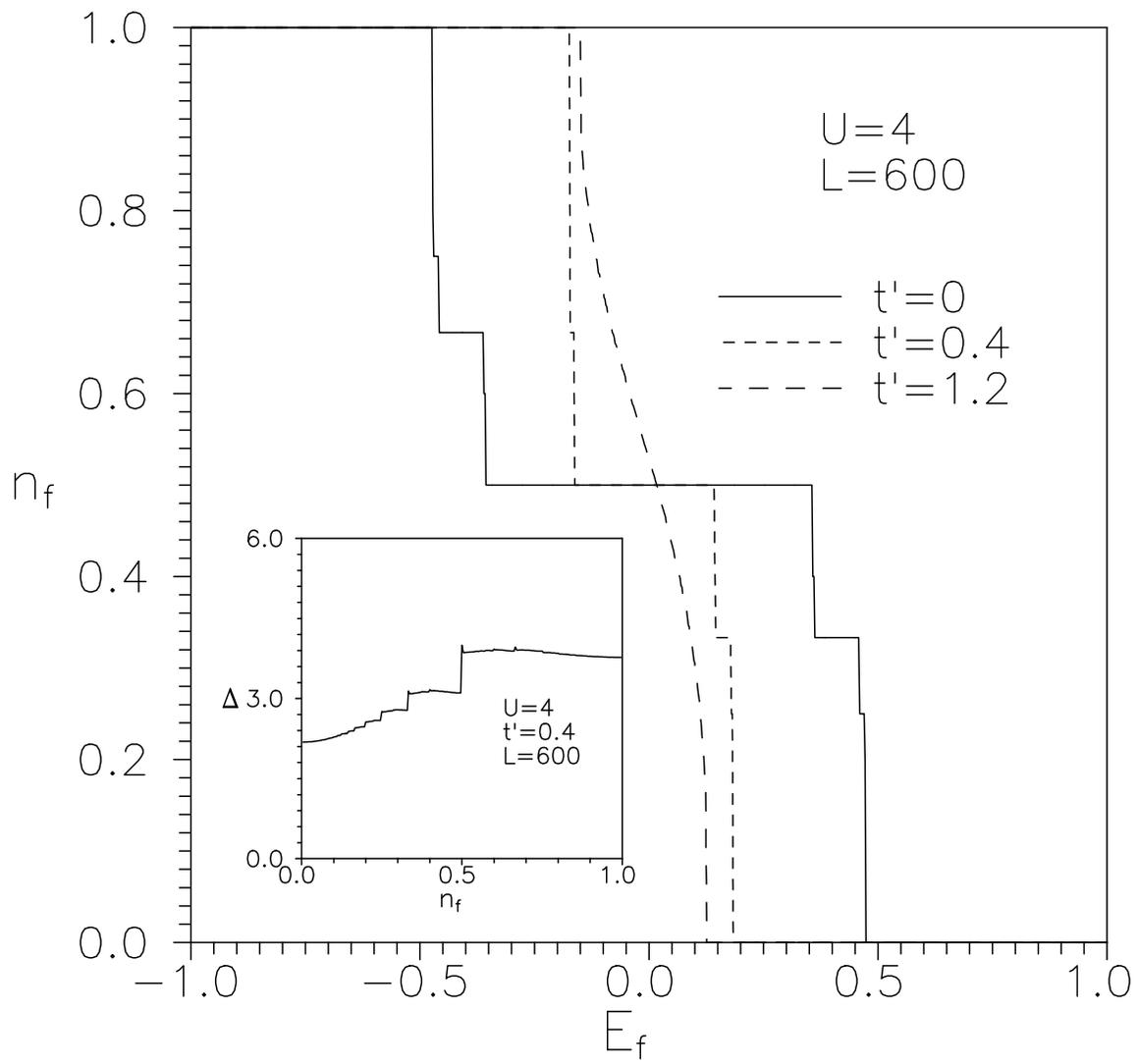
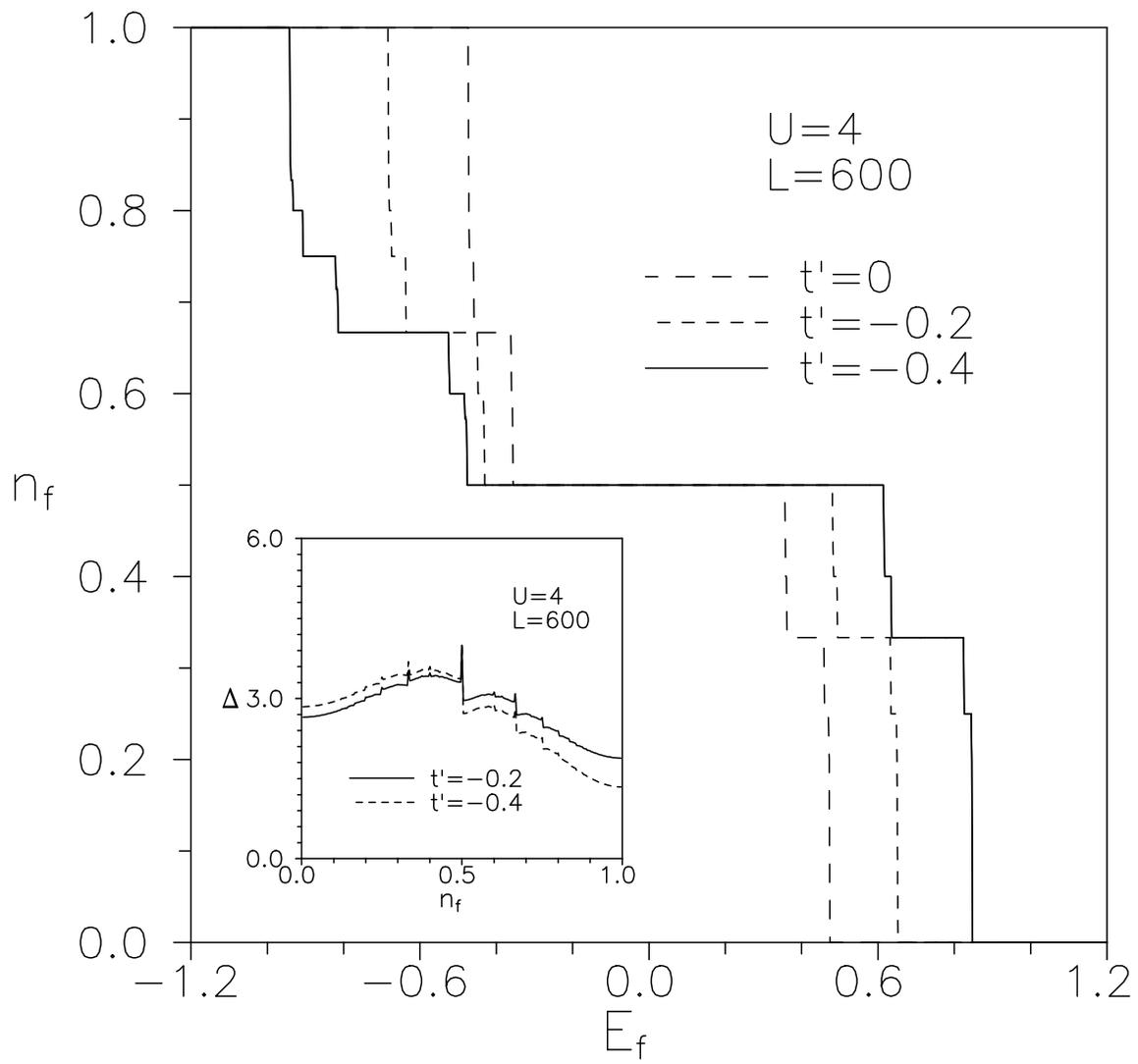
\end{document}